\documentclass[10pt,letterpaper,twocolumn]{article}
\usepackage{ol2}
\usepackage{cite}
\usepackage{times,mathptmx,amssymb,amsmath}
\usepackage{graphicx,color}

\newcommand{\eh}[1]{\ensuremath{\,\textrm{#1}}}
\begin{document}
\twocolumn[ 

\title{Self-trapping threshold in disordered nonlinear photonic lattices}

\author{U. Naether,$^{1}$
		M. Heinrich,$^{2,3}$
		Y. Lahini,$^{4}$
		S. Nolte,$^{3}$
		R.A. Vicencio,$^{1}$
		M.I. Molina,$^{1}$
		and A. Szameit$^{3}$}

\address{
$^1$Dpto. de F\'isica and MSI-Nucleus on Advanced Optics, Center for
	Optics and Photonics (CEFOP),\\
	Facultad de Ciencias, Universidad de Chile, Santiago, Chile\\
$^2$CREOL, The College of Optics and Photonics, University of Central Florida,
	Orlando, FL-32816, USA\\
$^3$Institute of Applied Physics, Abbe Center of Photonics, Friedrich-Schiller-Universit\"at Jena,
    Max-Wien-Platz 1, 07743 Jena, Germany\\
$^4$Department of Physics, 
MIT - Massachusetts Institute of Technology Cambridge, MA 02139, USA}

\begin{abstract}
	\noindent We investigate numerically and experimentally the
	influence of coupling disorder on the self-trapping dynamics in nonlinear
	one-dimensional optical waveguide arrays. The existence of a lower and
	upper bound of the effective average propagation constant 
	allows for a generalized definition of the threshold power for the onset of soliton
	localization. When compared to perfectly ordered systems, this threshold is
	found to decrease in the presence of coupling disorder.
\end{abstract}
]

\noindent
Discrete solitons \cite{rep1,rep2} emerge in periodic nonlinear lattices when self-focusing becomes sufficiently strong to balance diffraction. In spite of this very general
mechanism, the transition dynamics between extended and localized states depend
strongly on the geometry and dimensionality of the specific system. For example,
the critical self-trapping nonlinearity in a chain of coupled oscillators
decreases with the length of the chain, as can seen comparing the values of a
coupled dimer \cite{kenkre86}, smaller one-dimensional chains~\cite{mario93} and
very long ones~\cite{magnus95}. A first attempt to find a more general
criterion led to the discovery of a common ratio of $\approx 1.3$ between the
critical nonlinearity and the minimum bound-state energy for defect states in
lattices with different dimensions and topologies \cite{carlos00}. Experimentally, the transition was observed in arrays of AlGaAs \cite{meier}, in arrays made with the
fs-laser direct-writing technology \cite{arrays} and nematic liquid crystal arrays \cite{assanto}.

Disorder can manifest itself as random change to the local properties
of all lattice sites. Known as Anderson localization, the ensuing interference between multiple scattering paths
may significantly impede and even suppress the transverse transport of waves
\cite{al}. Exponential localization was first directly observed in disordered photonic lattices \cite{exp2d,exp1d}.
Similarly, random modifications of the coupling between lattice sites can lead to exponentially localized states \cite{nddo}, as was demonstrated in
waveguide arrays \cite{and} where coupling disorder corresponds to
variations of the waveguide separation \cite{coupl}. It is known that
Anderson localization can be destroyed by weak nonlinearity,
giving rise to unlimited subdiffusive spreading for very large systems
\cite{disnl6, flach1,flach2}. On the other hand, when the system size is
smaller than the (linear) localization length, the interplay of diagonal disorder
and nonlinearity induces a smoothening in the distribution of a spreading
initial single-site excitation \cite{ddoe}. In the limit of large
nonlinearities, self-trapping of a large fraction of the initial wave packet
occurs irrespective of disorder \cite{disnl5, disnl7}.

The question arises as to how the simultaneous action of both nonlinearity and
disorder influences not only the width of the output \cite{exp2d,exp1d, pertsch}, but especially the dynamics of self-trapping. The existence of a lower and upper bound of the average propagation constant was
shown recently for various ordered lattice configurations of different
dimensions, leading to the definition of a generalized power threshold
\cite{stt}.  In this letter, we show numerically and verify experimentally that the presence of coupling disorder entails a systematic decrease of this
threshold for dynamical excitation in one-dimensional optical waveguide
arrays.

In the coupled-mode approximation, the evolution of light in a 1D
lattice of $N$ waveguides along the propagation direction $z$ is described by a
discrete nonlinear Schr\"odinger-like equation:
%
\begin{equation}\label{dnls}
	i\frac{d u_n}{d z}+\varepsilon_nu_n+C_n u_{n+1}+C_{n-1}u_{n-1}+\gamma|u_{n}|^2 u_{n}=0\ ,
\end{equation}
%
where $u_{n}$ corresponds to the modal field amplitude at site $n$,
$C_{n}$ is the coupling between guides $n$ and $n+1$, and $\gamma=1$ ($\gamma=-1$) is the nonlinear coefficient for focusing (defocusing) nonlinearity. Hereafter, we take the nonlinear coefficient $\gamma=1$ and consider primarily the case without diagonal disorder, thus $\varepsilon_n=0$ . The effective strength
of the nonlinearity is then given by the conserved total power
(norm) $P\equiv\sum_{n} |u_{n}|^2$. Furthermore, we will consider single-site excitations,
i.e. $u_{n}(z=0)=\sqrt{P}\delta_{n,n_0}$, where $n_0$ corresponds to a central
position in the lattice. 
The second conserved quantity of the system, the Hamiltonian 
\begin{equation}\label{H}
	H\equiv  \sum_{n}
	\left[C_n(u_{n+1}^*u_{n}+u_{n}^*u_{n+1})+\frac{|u_n|^4}{2}\right] ,
\end{equation}
then has the value of $H=P^2/2$. In order to evaluate the size of the wave
packet, we use the participation number
\begin{equation}\label{R}
	R\equiv\frac{P^2}{\sum_{n}|u_{n}|^4} \,,
\end{equation}
which approaches $1$ for a highly localized wave packet and tends to $N$ for an extended excitation. $R$ serves as a measure of how many lattice sites
effectively contribute to a given profile. In comparison to other quantities
such as the variance or standard deviation, the participation number emphasizes 
substantially excited sites and is less sensitive to weak excitations that may propagate far away from the initial site.

In analogy to stationary solutions $u_{n}(z)=u_{n} \exp{(i
\lambda z)}$, we assign an effective propagation constant
$\lambda_e$ to any momentary profile. This quantity can be obtained by
multiplying Eq.\eqref{dnls} by $u_{n}^*$ and taking the sum 
over all lattice sites:
\begin{equation}\label{L}
	\lambda_e \sum_{n} |u_{n}|^2=\sum_{n} \left[C_n(u_{n+1}^*u_{n}+u_{n}^*u_{n+1})+ |u_{n}|^4\right]\nonumber.
\end{equation}
This expression is related to the Hamiltonian: 
\begin{equation}\label{ex1}
	\lambda_e P=H+\frac{1}{2} \sum_{n}
	|u_{n}|^4=H+\frac{1}{2}\frac{P^2}{R}.\nonumber
\end{equation}
After inserting $H= P^2/2$, we finally arrive at
\begin{equation}\label{ex1}
	\lambda_e =\frac{P}{2}\left(1+\frac{1}{R}\right).
\end{equation}
%
\begin{figure}[t]
\centering
\includegraphics[width=\linewidth]{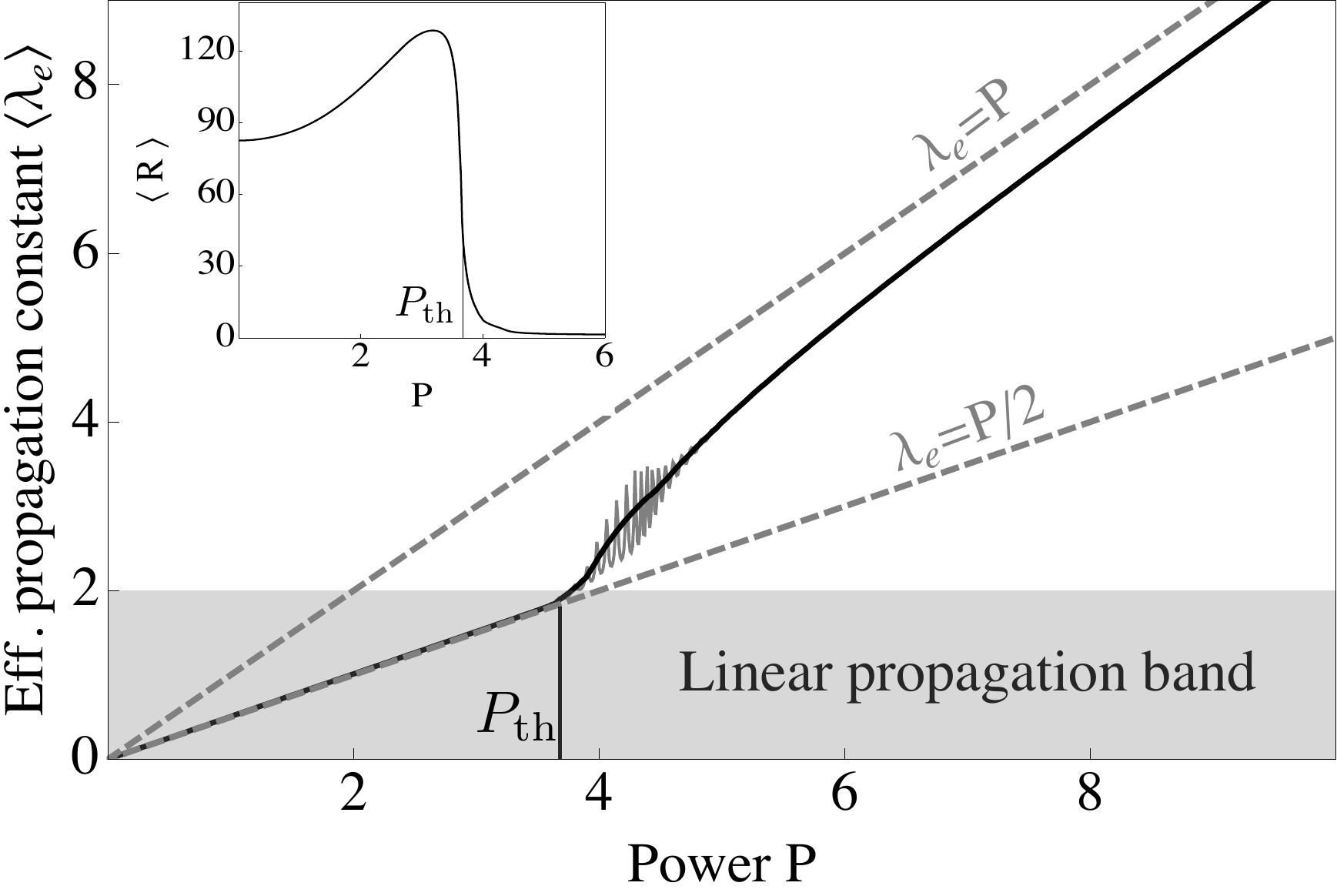}
\caption{Power-dependent effective
	propagation constant. Shown is the mean $ \langle \lambda_e \rangle $ (solid black graph) and  $ \lambda_e|_{z=N/4} $ (solid gray)
	obtained from central single-site excitations in a homogeneous lattice
	($C_n=1$) with $N=221$. Inset: $\langle R\rangle$ vs. P.} 
\label{fig1}
\end{figure}

\noindent
Let us first consider the limiting cases. In the strongly nonlinear
regime, diffraction is totally suppressed and the wave packet remains localized
with a participation ratio of $R\approx1$; the effective propagation
constant becomes equal to the power: $\lambda_e=P$. Conversely, in a
diffractive process, the wave packet spreads across the whole lattice. The value
of $R$ thus increases up to a value of the order of $N$, yielding $\lambda_e=P/2$ for sufficiently large systems.

Figure~\ref{fig1} illustrates the power-dependent behavior of $\lambda_e$
for a single-site initial excitation at the center of a uniform lattice with $N=221$
waveguides, the inset shows $\langle R\rangle$ vs.  $P$. At vanishing excitation power, the $\delta$-like input pattern
corresponds to a flat excitation of all modes of the lattice; $\lambda_{e}$ thus
constitutes the average over the propagation band and emerges at its center
($\lambda_{e}=0$). With increasing power, nonlinear contributions become
relevant as the modes start to interact and to exchange energy. Consequently,
$\lambda_{e}$ exhibits oscillations along $z$ with a power-dependent period.
It has been shown \cite{stt} that $\lambda_e$ represents the
average propagation constant of all excited modes. Its
behavior inside the sector $P/2 \leqslant
\lambda_e \leqslant P$ serves as indicator as to whether the wave packet's evolution is dominated by
diffraction or localization. The power-dependent transition between those two
regimes becomes clearly visible when the rapid oscillations of $\lambda_e(P)$
are removed by averaging over a certain interval of propagation distances (here:
$z\in[N/10,N/4]$). In the following, we define the threshold power
$P_\textrm{th}$ as the lowest value where $\lambda_e$ exceeds the linear limit
of $P/2$ by a certain fixed cutoff value $\Delta$.
\begin{figure}[b]
\centering
\includegraphics[width=\linewidth]{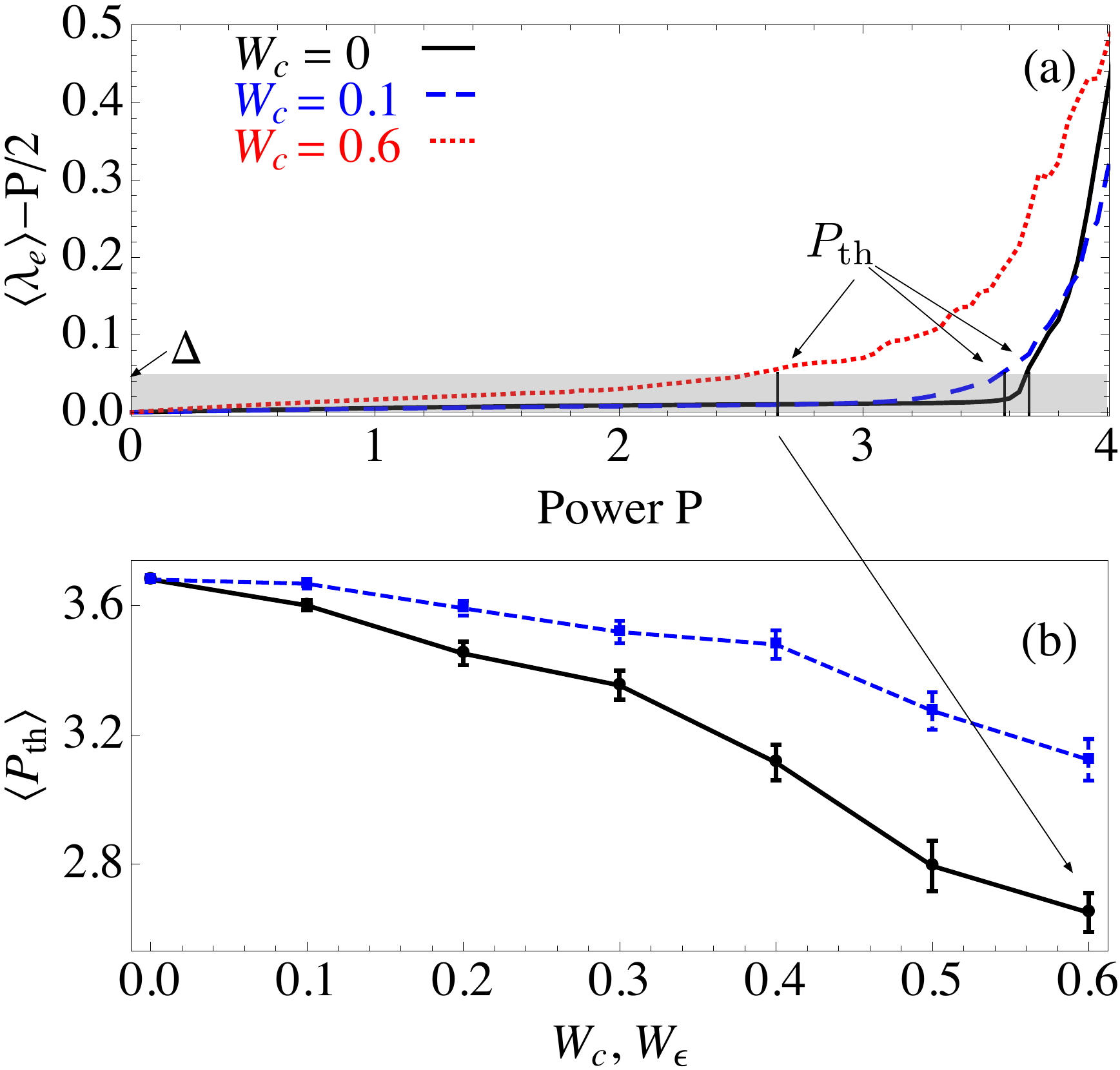}
\caption{(Color online) (a) Transition between linear behavior
	and nonlinear localization for varying off-diagonal disorder. Shown is
	$\langle\lambda_e\rangle-P/2$ for $W_c=0$ ( solid black),
	$W_c=0.1$ (dashed blue) and $W_c=0.6$ (dotted red). (b) Ensemble averages and
	standard error of $P_\textrm{th}$ obtained for $\Delta=0.05$ for off- (on-)diagonal disorder with full (dashed) lines. }
\label{fig2}
\end{figure}

We now turn to lattices with coupling disorder, where the individual coupling
constants $C_n=1+\mu$ are uniformly distributed with $\mu\in[-W_{c}/2,W_{c}/2]$. 
Sets of $50$ realizations each for different degrees of disorder were independently evaluated. Figure~\ref{fig2}(a) shows the ensemble-averaged value of
$\langle\lambda_e\rangle-P/2$ for off-disorder. For small $P$, all curves remain close to zero. 
Note, however, that the most strongly disordered ensemble (with $W=0.6$)  exhibits the largest
values. This illustrates the influence of Anderson localization, which gives
rise to less extended linear modes. Consequently the condition $R\gg1$ is
fulfilled to a lesser degree, for increasing disorder. The decrease for on-diagonal disorder is weaker.  Due to the lack of a threshold for the existence of nonlinear stationary modes in 1D lattices, the onset of localization is gradual \cite{stt}. Nevertheless, a pronounced transition is
visible; a choice of $\Delta=0.05$ provides a good measure for the departure from the linear behavior.
Clearly, disorder serves to smoothen the transition in the ensemble averages, but also systematically lowers the power threshold for the onset of
nonlinear localization [see black curve in Fig.~\ref{fig2}(b)]. For comparison, the threshold for equivalent amounts of on-diagonal disorder is shown [dashed curve in Fig.~\ref{fig2}(b)].

\begin{figure}[t]
\centering
\includegraphics[width=\linewidth]{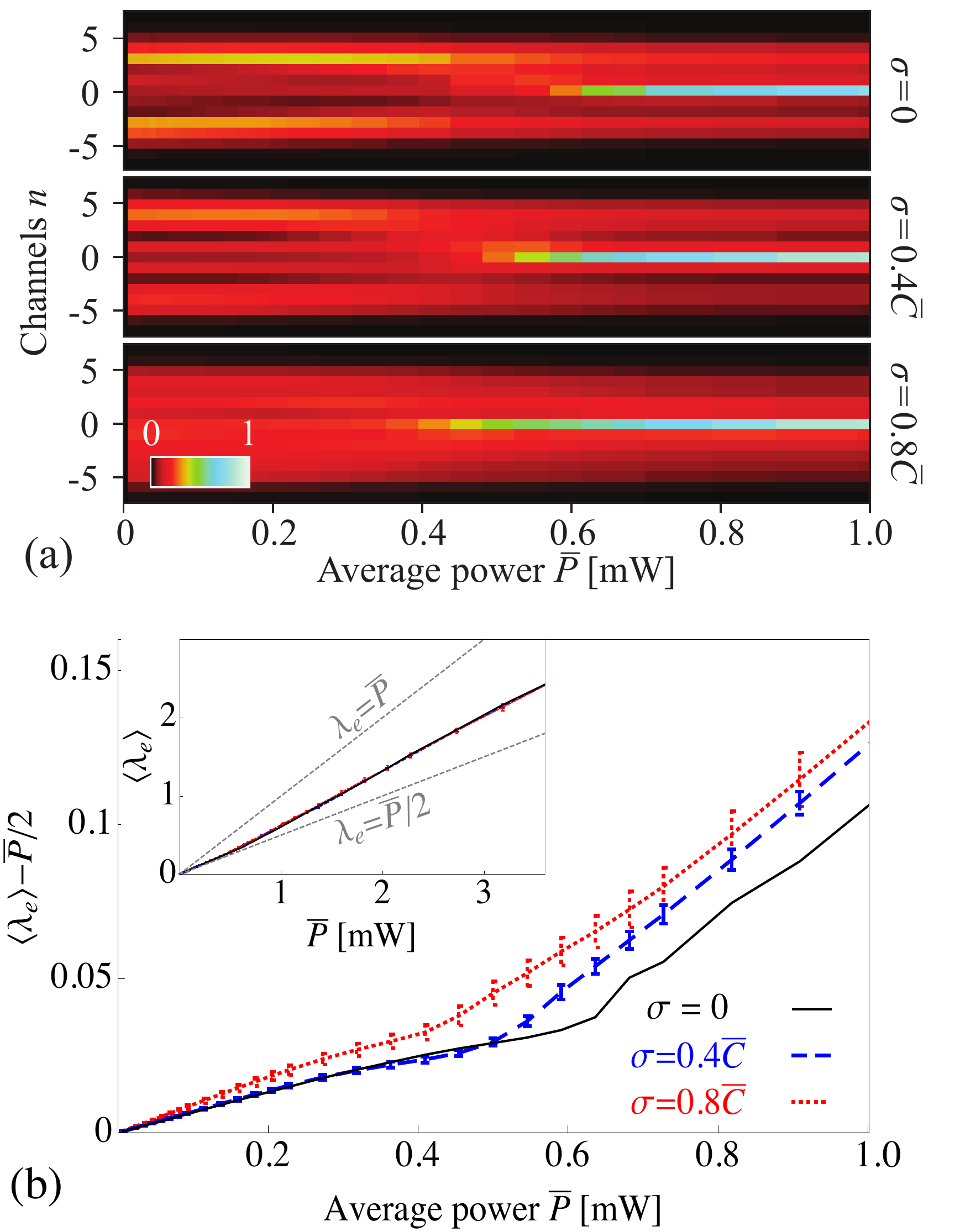}
\caption{(Color online) (a) Experimental mean output intensity distributions 
		 at $z=1.28\pi/2\bar{C}$ for $\sigma=0\ \bar{C}$ [top], $\sigma=0.4 \ \bar{C}$ [center]
		 and $\sigma=0.8\ \bar{C}$ [bottom].
		 (b) Transition region: Power-dependent deviation from linear behavior
		 $\langle\lambda_e\rangle-\bar{P}/2$ for $\sigma=0\ \bar{C}$ (solid
		 black), $\sigma=0.4\ \bar{C}$ (dashed blue) and $\sigma=0.8\ \bar{C}$
		 (dotted red).
		 Inset: The effective propagation constant remains confined to the
		 interval $P/2 \leqslant \lambda_e \leqslant P$.
 		}
\label{fig3}
\end{figure}

To verify our results experimentally, we fabricated waveguide arrays in a
$100\eh{mm}$ long fused silica sample using the fs-laser direct-writing
technology \cite{arrays}. The coupling disorder $C=\bar{C}\pm\sigma$ was
realized for three different values ($\sigma=0$, $0.4\,\bar{C}$ and $0.8\,\bar{C}$) by
varying the waveguide separations \cite{coupl} and ensembles of 18 independent realizations were examined. The sample length corresponded to a propagation distance of
$1.28\pi/2\bar{C}$ with $\bar{C}=0.02\eh{mm}^{-1}$. Nonlinear excitation
was achieved by  a Ti:sapphire laser system, delivering $300\eh{fs}$
pulses with a repetition rate of $1\eh{kHz}$ at $800\eh{nm}$. 
The plots in Fig.~\ref{fig3}(a) show the ensemble-averaged output intensity
distributions obtained with average input powers between $0
<\bar{P}\le1.0\eh{mW}$ for vanishing, intermediate and high disorder.
Despite the diffractive background from the pulsed excitation, the
accelerated onset of localization is clearly visible for the disordered cases.
Figure~\ref{fig3}(b) substantiates this observation by means of the effective
propagation constant obtained from the patterns. In agreement with our
simulations, the dynamics are generally confined to the region $P/2 \leqslant
\lambda_e \leqslant P$ [see inset]. Note that due to the limited propagation
length, $\lambda_e$ systematically exceeds $P/2$. Nevertheless, the
transition to the regime of nonlinear localization is clearly visible in all
cases, and is shifted towards lower $\bar{P}$ for larger degrees of disorder
$\sigma$.

In conclusion, we have analyzed the nonlinear localization dynamics in
disordered photonic lattices. Based on the definition of a dynamical
threshold that relies on analytical estimates of the upper and lower
bounds of the propagation constant, we have shown numerically and verified
experimentally that coupling disorder may strongly inhibit transport, causing fewer sites to be substantially excited and systematically decreases the threshold
power required to dynamically excite nonlinear localized wave packets. 

The authors gratefully acknowledge funding  by FONDECYT Grants 1110142, 1120123,
a CONICYT doctoral fellowship, Programa ICM P10-030-F and Programa de Financiamiento Basal de CONICYT (FB0824/2008). M.H. was supported by the German
National Academy of Sciences Leopoldina (grant No. LPDS 2012-01). A.S. thanks the German Ministry of Education and Research for financial support (ZIK 03Z1HN31).


\vspace{-0.5cm} 
 
\begin{thebibliography}{99}
\bibitem{rep1}S. Flach and A. Gorbach, Phys. Rep. {\bf 467}, 1 (2008).
\bibitem{rep2}F. Lederer, G.I. Stegeman, D.N. Christodoulides, G. Assanto, M. Segev, and Y. Silberberg, Phys. Rep. \textbf{463}, 1 (2008).
\bibitem{kenkre86}V.M. Kenkre and D.K. Campbell, \prb \textbf{34}, 4959 (1986).
\bibitem{mario93}M.I. Molina and G.P. Tsironis, Physica D \textbf{65}, 267 (1993).
\bibitem{magnus95}M. Johansson, M. H\"ornquist, and R. Riklund, \prb \textbf{52}, 231 (1995).
\bibitem{carlos00}C.A. Bustamante and M.I. Molina, \prb \textbf{62}, 15287
(2000).
\bibitem{meier} J. Meier, J. Hudock, D.N. Christodoulides, G.I. Stegeman, H.Y. Yang, G. Salamo, R. Morandotti, J.S. Aitchison, and Y. Silberberg, J. Opt. Soc. Am. B {\bf 22}, 1432 (2005).
\bibitem{arrays} A. Szameit, D. Bl\"omer, J. Burghoff, T. Schreiber, T. Pertsch, S. Nolte, and A. T\"unnermann, Opt. Exp. \textbf{13}, 10552 (2005).
\bibitem{assanto}A. Fratalocchi, G. Assanto, K. Brzdakiewicz, and M.  Karpierz, Opt. Exp. {\bf 13}, 1808 (2005).
\bibitem{al}P.W. Anderson, Phys. Rev. {\bf 109}, 1492 (1958).
\bibitem{exp2d} T. Schwartz, G. Bartal, S. Fishman, M. Segev,  Nature {\bf 446}, 52 (2007).
\bibitem{exp1d} Y. Lahini, A. Avidan, F. Pozzi, M. Sorel, R. Morandotti, D.N. Christodoulides, Y. Silberberg,  Phys. Rev. Lett.{\bf 100}, 013906 (2008).
\bibitem{nddo} C.M. Soukoulis and E.N. Economou,  Phys. Rev. B {\bf 24}, 5698 (1981).
\bibitem{and}  A. Szameit, Y.V. Kartashov, P. Zeil, F. Dreisow, M. Heinrich, R.
Keil, S. Nolte and A. T\"{u}nnermann, Opt. Lett. {\bf 35}, 1172 (2010).
\bibitem{coupl} A. Szameit, F. Dreisow, T. Pertsch, S. Nolte, and A. T\"unnermann, Opt. Exp. \textbf{15}, 1579 (2006).
\bibitem{disnl6}A.S. Pikovsky and D.L. Shepelyansky, Phys. Rev. Lett. {\bf 100}, 094101 (2008).
\bibitem{flach1}M.V. Ivanchenko, T.V. Laptyeva, and S. Flach, \prl {\bf 107}, 240602 (2011).
\bibitem{flach2}J.D. Bodyfelt, T.V. Laptyeva, G. Gligoric, D.O. Krimer, Ch. Skokos, S. Flach, Int. J. Bifurcat. Chaos {\bf 21}, 2107 (2011)
\bibitem{ddoe} U. Naether, S. Rojas-Rojas, A.J. Mart\'inez, S. St\"utzer, A.T\"{u}nnermann, S. Nolte, M. I. Molina, R.A. Vicencio, and A. Szameit,  Opt.Exp. {\bf 21}, 927 (2013).
\bibitem{disnl5}G. Kopidakis, S. Komineas, S. Flach, and S. Aubry, Phys. Rev. Lett. {\bf 100}, 084103 (2008).
\bibitem{disnl7}S. Flach, D. O. Krimer, and Ch. Skokos, Phys. Rev. Lett. {\bf 102}, 024101 (2009).
\bibitem{pertsch} T. Pertsch, U. Peschel, J. Kobelke, K. Schuster, H. Bartelt, S. Nolte, A. T\"unnermann, and Falk Lederer,  \prl {\bf 93}, 053901 (2004).
\bibitem{stt}U. Naether, A. J. Mart\'inez, D. Guzm\'an-Silva, M.I. Molina, R.A.Vicencio, submitted (2012).



\end{thebibliography}
\end{document}